\documentclass[twocolumn,superscriptaddress,longbibliography]{revtex4-1} 

\usepackage{graphicx} 
\usepackage{natbib} 
\usepackage[usenames,dvipsnames]{color} 
\usepackage{amsmath,amssymb,amstext,dsfont} 
\usepackage{amssymb} 
\usepackage{soul} 
\usepackage{ifthen} 

\newcommand{\nn}{\nonumber}

\newcommand{\bra}[1]{\langle{#1}|}
\newcommand{\ket}[1]{|{#1}\rangle}

\def\l{\left}
\def\r{\right}
\newcommand{\sch}{Schr\"odinger~}
\def\be#1\ee{\begin{equation}#1\end{equation}}
\def\ba#1\ea{\begin{align}#1\end{align}}
\def\bg#1\eg{\begin{gather}#1\end{gather}}
\def\t{\text}

\def\eq#1{(\ref{eq:#1})}

\def\shownote{1} 
\newcommand{\note}[1]{\ifthenelse{\shownote=1}{\textcolor{Red}{[[#1]]}}{.}}
\def\showaddmat{1} 
\newcommand{\addmat}[1]{\ifthenelse{\showaddmat=1}{\textcolor{Gray}{[[#1]]}}{}}

\begin{document}

\title{Dynamics of a two-level system under strong driving: quantum gate optimization based on Floquet theory}

\author{Chunqing Deng}
\altaffiliation{Current address: D-Wave Systems Inc., Burnaby, BC, Canada V5G 4M9}
\email[Electronic address: ]{dengchunqing@gmail.com}
\affiliation{Institute for Quantum Computing, Department of Physics and Astronomy, and Waterloo Institute for Nanotechnology, University of Waterloo, Waterloo, ON, Canada N2L 3G1}

\author{Feiruo Shen}
\affiliation{Institute for Quantum Computing, Department of Physics and Astronomy, and Waterloo Institute for Nanotechnology, University of Waterloo, Waterloo, ON, Canada N2L 3G1}

\author{Sahel Ashhab}
\affiliation{Qatar Environment and Energy Research Institute, Hamad Bin Khalifa University, Qatar Foundation, Doha, Qatar}

\author{Adrian Lupascu}
\affiliation{Institute for Quantum Computing, Department of Physics and Astronomy, and Waterloo Institute for Nanotechnology, University of Waterloo, Waterloo, ON, Canada N2L 3G1}

\date{\today}

\begin{abstract}
We consider the dynamics of a two-level system (qubit) driven by strong and short resonant pulses in the framework of Floquet theory. First we derive analytical expressions for the quasienergies and Floquet states of the driven system. If the pulse amplitude varies very slowly, the system adiabatically follows the instantaneous Floquet states, which acquire dynamical phases that depend on the evolution of the quasienergies over time. The difference between the phases acquired by the two Floquet states corresponds to a qubit state rotation, generalizing the notion of Rabi oscillations to the case of large driving amplitudes. If the pulse amplitude changes very fast, the evolution is non-adiabatic, with transitions taking place between the Floquet states. We quantify and analyze the nonadiabatic transitions during the pulse by employing adiabatic perturbation theory and exact numerical simulations. We find that, for certain combinations of pulse rise and fall times and maximum driving amplitude, a destructive interference effect leads to a remarkably strong suppression of transitions between the Floquet states. This effect provides the basis of a quantum control protocol, which we name Floquet Interference Efficient Suppression of Transitions in the Adiabatic basis (FIESTA), that can be used to design ultra-fast high-fidelity single-qubit quantum gates.
\end{abstract}

\maketitle

\section{Introduction}

Fault-tolerant quantum computing requires that the fidelities of elementary quantum logic gates exceed certain thresholds~\cite{Nielsen:2011:QCQ:1972505}. These fidelity thresholds depend on the architecture. However, even for the most forgiving cases the threshold is high; therefore, reaching and exceeding it is an extraordinarily difficult task. The implementation of high-fidelity quantum operations is also relevant, although to a lesser extent, in other quantum technologies, including quantum communication and sensing. Qualitatively speaking, the fidelity of a gate depends on the ratio between the quantum coherence time and the gate time, implying two approaches to increase gate fidelities: increasing coherence times and reducing gate times.

In most qubit architectures, single-qubit gates are implemented by applying an oscillating field that is resonant with the qubit transition frequency. Most of the experiments and theoretical studies on such gates were performed with a driving strength that is significantly smaller than the qubit transition frequency. In this weak-driving regime, the qubit undergoes Rabi oscillations between the two energy eigenstates at a rate that is proportional to the driving strength. This situation is described rather accurately using the rotating wave approximation~\cite{Rabi} (RWA), which predicts sinusoidal oscillations between the energy eigenstates.

With strong driving, the RWA breaks down and the dynamics becomes very complex.
Despite the increased complexity in controlling dynamics, the strong driving regime is interesting because it brings the opportunity to implement advanced quantum control~\cite{Warren1581, Alessandro, Khaneja2001, BrumerShapiro200303, vandersypen_2004_revNMR} and achieve faster quantum operations than in the weak driving regime.

A two-level quantum system (or a qubit) driven by strong resonant pulses has been studied experimentally using NV centers in diamond~\cite{Fuchs:2009ca, scheuer_precise_2014}, semiconductor quantum dots~\cite{Petta669, gaudreau2012coherent, Stehlik2012} and superconducting circuits~\cite{Nakamura:2001cr, Chiorescu2004, saito_2004_multiphoton,Oliver:2005rK, Sillanpaeae_2006_LZCPB, Saito2006, Wilson:2007dd, izmalkov_2008_coupledqubits, sun_2009_LZ, Tuorila2010, Silveri:2014t_, Yoshihara:2014db, Deng2015, shytov_2003_interferometry, Ashhab:2007ep, Shevchenko:2010hf}. In a few experiments~\cite{Fuchs:2009ca, Chiorescu2004, Deng2015}, the qubit population exhibits complex dynamics containing a few frequency components, as the driving strength approaches or exceeds the qubit transition frequency, a signature of the breakdown of the RWA. Therefore, for the proper design of quantum gates in this regime, different theoretical methods are required. Floquet theory~\cite{Shirley:1965sI}, which provides a general framework for treating periodically driven quantum systems with any driving strength, is the natural method for analyzing strong-driving dynamics. In Ref.~\onlinecite{Deng2015}, the quasienergies and quasienergy states (hereafter referred to as Floquet states), as predicted by Floquet theory, were observed in a strongly driven superconducting qubit. Moreover, the observed dynamics pointed to the important role that the pulse shape plays in the qubit evolution, as determined by adiabaticity conditions in the Floquet picture.

In this paper, we perform a comprehensive theoretical analysis of qubit dynamics under strong resonant pulses using Floquet theory. We derive approximate analytical expressions for the quasienergies and Floquet states as functions of the driving amplitude for the practically important case where the qubit is biased at its symmetry point. We then analyze the qubit dynamics induced by a driving pulse, obtained by modulating the amplitude of a periodic signal. The quantum state of the qubit is naturally expressed as a superposition of Floquet states. The occupation probabilities of the Floquet states remain fixed as long as the driving strength is fixed, but they can change when the pulse envelope varies in time. We analyze the qubit dynamics using a representation of its state in the Floquet picture~\cite{Drese:1999tq, Guerin2003}. When the change in the driving strength is slow, the occupation probabilities of the Floquet states remain almost constant, although the Floquet states themselves change in accordance with the instantaneous driving strength. The quantum superposition of the Floquet states acquires a phase that depends on the evolution of the quasienergies over time. This acquired phase corresponds to a qubit state rotation, generalizing the notion of Rabi oscillations to the case of large driving amplitudes. However, a pulse with slowly varying amplitude corresponds to relatively slow quantum gates. We therefore analyze the dynamics with short pulses, when in general nonadiabatic transitions between Floquet states cannot be neglected. We find that with suitable pulse parameters the nonadiabatic transitions can be largely suppressed. This effect provides the basis of Floquet Interference Efficient Suppression of Transitions in the Adiabatic basis (FIESTA), a method to optimize quantum gates with strong driving. Finally, we show that FIESTA can be used to implement high-fidelity single-qubit operations in very short times, significantly alleviating the effect of decoherence.

The remainder of this paper is organized as follows: In Sec.~\ref{sec:Floquet}, we derive expressions for the quasienergies and Floquet states of a qubit biased at the symmetry point under harmonic driving with arbitrary strength. In Sec.~\ref{sec:adiabatic_theory}, we describe the adiabatic theory in the Floquet picture and derive the adiabatic condition. In Sec.~\ref{sec:adiabatic}, we use the adiabatic theory in the Floquet picture to describe the quantum state evolution effected by pulses with slowly varying amplitude. In Sec.~\ref{sec:nonadiabatic}, we present a quantitative analysis of the nonadiabatic transitions between the Floquet states when the amplitude of the pulse changes rapidly. In Sec.~\ref{sec:gates}, we calculate the fidelities of single-qubit gates optimized so as to suppress nonadiabatic transitions. In Sec.~\ref{sec:OQC}, we discuss the connections between FIESTA and other optimal quantum control methods. Section~\ref{sec:conclusion} contains concluding remarks.

\section{Floquet theory for a driven qubit} \label{sec:Floquet}

In this section we discuss the quasienergies and Floquet states of a qubit driven with a single-frequency tone. Some of the results in this section have been presented in our previous paper~\cite{Deng2015} and are included here for completeness. We start with a general model in which a qubit is driven by a harmonic field. The Hamiltonian is given by
\begin{equation} \label{eq:Hamstart}
H = - \frac{\Delta}{2} \sigma_z + A \cos \left( \omega t +\phi \right) \sigma_x,
\end{equation}
with $\sigma_\alpha$ ($\alpha= x,y,z$) the Pauli matrices. The ground ($\ket{0}$) and excited ($\ket{1}$) states are ordered such that $\sigma_z\ket{0}=\ket{0}$ and $\sigma_z\ket{1}=-\ket{1}$. The Hamiltonian in Eq.~(\ref{eq:Hamstart}) is written under the standard convention where in the absence of driving the energy eigenstates are the ground and excited states $\ket{0}$ and $\ket{1}$. In a frame rotating about the $z$ axis with frequency $\omega$, the Hamiltonian is:
\ba \label{eq:HamRF}
H_{\t{RF}} = &-\frac{\Delta-\omega}{2}\sigma_z \nn \\
&+ A \cos(\omega t + \phi) \l(\cos(\omega t)\sigma_x + \sin(\omega t)\sigma_y \r).
\ea
In the weak driving limit ($A\ll\omega$), the terms in Eq.~\eq{HamRF} oscillating at frequencies $2\omega$ can be ignored under RWA. With this approximation the Hamiltonian is given by
\be \label{eq:HRWA}
H_{\t{RWA}} = -\frac{\Delta-\omega}{2}\sigma_z + \frac{A}{2}(\cos\phi\,\sigma_x + \sin\phi\,\sigma_y).
\ee
This is a well-known time-independent Hamiltonian representing a spin-1/2 particle precessing in a magnetic field. In this paper, we will analyze the dynamics under the more general Hamiltonian, Eq.~\eq{Hamstart}, for large driving amplitude. We will refer to Eq.~\eq{HRWA} while connecting the general driven dynamics to the special cases in the well-known weak driving regime. For simplicity, we will treat Eq.~\eq{Hamstart} under $\phi = 0$ in the remainder of this section. Our derivation can be easily generalized to arbitrary $\phi$.

Another alternative form of Eq.~\eq{Hamstart}, which is related to it by a $\pi/2$ rotation about the $y$ axis, is
\begin{equation} \label{eq:Hamrotated}
H_{\rm rot} = - \frac{\Delta}{2} \sigma_x - A \cos \left( \omega t \right) \sigma_z.
\end{equation}
This form is convenient for our analytical derivations for the quasienergies and Floquet states (in which we will closely follow Ref.~\onlinecite{Son:2009eg}), and we shall therefore use this form of the Hamiltonian in the remainder of this section.

According to Floquet theory, for a periodic Hamiltonian there exist solutions to the Schr\"odinger equation of the form
\begin{equation} \label{eq:generalfloquet}
\ket{\psi_{F,j}(t)} = e^{-i\epsilon_jt} \ket{u_j(t)},
\end{equation}
where $\epsilon_j$ are the quasienergies and $\ket{u_j(t)}$, to which we shall refer as Floquet modes, are periodic with the periodicity of the Hamiltonian. Because of the periodicity, we can write
\begin{equation}
\ket{u_j(t)} = \sum_{n=-\infty}^{\infty} e^{in\omega t} \ket{u_{j,n}},
\label{eq:BasicFormOfFloquetStates}
\end{equation}
with $\omega=2\pi/T$, where $T$ is the period of the Hamiltonian, and the state Fourier components $\ket{u_{j,n}}$ are vectors in the Hilbert space of the system. Each Fourier coefficient $\ket{u_{j,n}}$ contains two complex numbers corresponding to the states $\ket{0}$ and $\ket{1}$: $ \ket{u_{j,n}} = \left( u_{j,n,0} \ket{0} + u_{j,n,1} \ket{1} \right)$. The index $j$ in Eq.~(\ref{eq:generalfloquet}) takes a number of values equal to the dimension of the Hilbert space. Importantly, the Floquet states form a complete basis, such that any quantum state can be expressed as a superposition of Floquet states. With a fixed driving strength, the solution of the \sch equation is a superposition of Floquet states with fixed coefficients.

Substituting Eqs.~\eq{generalfloquet} and \eq{BasicFormOfFloquetStates} in the time-dependent \sch equation for a qubit with the Hamiltonian given by Eq.~(\ref{eq:Hamrotated}), we find the relation
\ba
\epsilon_j \ket{u_{j,n}} = &\left( - \frac{\Delta}{2} \sigma_x + n\omega \right) \ket{u_{j,n}} \nn \\
&- \frac{A}{2} \sigma_z \left( \ket{u_{j,n-1}} + \ket{u_{j,n+1}} \right).
\ea
The above set of equations can be expressed as a single equation:
\begin{equation}
\epsilon_j \ket{U_j} = H_F \ket{U_j}
\end{equation}
where $U_j$ is the vector $\{\dots , u_{j,n-1,0} , u_{j,n-1,1} , u_{j,n,0} , u_{j,n,1} , u_{j,n+1,0} , u_{j,n+1,1} , \dots \}$, and
\begin{widetext}
\begin{equation}
H_F = \left(
\begin{array}{cccccccc}
\ddots & & & & & & & \\
& (n-1) \omega & -\frac{\Delta}{2} & -\frac{A}{2} & 0 & 0 & 0 & \\
& {-\frac{\Delta}{2}} & (n-1) \omega & 0 & {\frac{A}{2}} & 0 & 0 & \\
& {-\frac{A}{2}} & 0 & n \omega & {-\frac{\Delta}{2}} & {-\frac{A}{2}} & 0 & \\
& 0 & {\frac{A}{2}} & {-\frac{\Delta}{2}} & n \omega & 0 & {\frac{A}{2}} & \\
& 0 & 0 & {-\frac{A}{2}} & 0 & (n+1) \omega & {-\frac{\Delta}{2}} & \\
& 0 & 0 & 0 & {\frac{A}{2}} & {-\frac{\Delta}{2}} & (n+1) \omega & \\
& & & & & & & \ddots
\end{array}
\right),
\label{eq:FloquetHamiltonian}
\end{equation}
\end{widetext}
is known as the Floquet Hamiltonian.

To solve for the eigenvalues and eigenstates of $H_F$, we perform a basis transformation to a rotating frame with a time-dependent rotation frequency and truncate the transformed Floquet Hamiltonian to a $2\times 2$ matrix. The detailed derivation can be found in Appendix~\ref{appendix}. We obtain the eigenvalues (i.e. the quasienergies):
\ba
\epsilon_0 & = \frac{1}{2} \left( - \omega - \sqrt{ \left[ \omega - \Delta J_0\left(\frac{2A}{\omega}\right) \right]^2 + \Delta^2 J_1^2\left(\frac{2A}{\omega}\right) } \right), \nonumber \\
\epsilon_1 & = \frac{1}{2} \left( - \omega + \sqrt{ \left[ \omega - \Delta J_0\left(\frac{2A}{\omega}\right) \right]^2 + \Delta^2 J_1^2\left(\frac{2A}{\omega}\right) } \right). \label{eq:QuasienergiesAnalytical}
\ea
In the basis of Eq.~\eq{Hamrotated}, the eigenvectors are given by:
\ba
\ket{u_{0,n}} &= \frac{1}{\sqrt{2}}\left( \begin{array}{c} {\cos\frac{\theta}{2}J_{n+1}\left(\frac{A}{\omega}\right) + \sin\frac{\theta}{2}J_{n}\left(\frac{A}{\omega}\right)} \\ \\ {-\cos\frac{\theta}{2}J_{n+1}\left(-\frac{A}{\omega}\right) + \sin\frac{\theta}{2}J_{n}\left(-\frac{A}{\omega}\right)} \end{array} \right), \nonumber \\
\ket{u_{1,n}} &= \frac{1}{\sqrt{2}}\left( \begin{array}{c} {-\sin\frac{\theta}{2}J_{n+1}\left(\frac{A}{\omega}\right) + \cos\frac{\theta}{2}J_{n}\left(\frac{A}{\omega}\right)} \\ \\ {\sin\frac{\theta}{2}J_{n+1}\left(-\frac{A}{\omega}\right) + \cos\frac{\theta}{2}J_{n}\left(-\frac{A}{\omega}\right)} \end{array} \right), \label{eq:PeriodicPartsOfFloquetStatesOriginalBasis}
\ea
with
\begin{equation}
\tan\theta = \frac{\Delta J_1\left(\frac{2A}{\omega}\right)}{\omega - \Delta J_0\left(\frac{2A}{\omega}\right)}. \label{eq:theta}
\end{equation}

The Rabi frequency is given by the difference between the two quasienergies (see Eq.~\eq{QuasienergiesAnalytical}):
\begin{equation}
\Omega_R = \sqrt{ \left[ \omega - \Delta J_0\left(\frac{2A}{\omega}\right) \right]^2 + \Delta^2 J_1^2\left(\frac{2A}{\omega}\right) }.
\label{eq:RabiFrequencyAnalyticalFormula}
\end{equation}
In the case of exact resonance (i.e.~$\omega=\Delta$), we obtain the expression
\begin{equation}
\Omega_R = \omega \sqrt{ \left[ 1 - J_0\left(\frac{2A}{\omega}\right) \right]^2 + J_1^2\left(\frac{2A}{\omega}\right) }.
\label{eq:RabiFrequencyAnalyticalFormulaResonantCase}
\end{equation}
This expression reduces to the well-known expressions in the weak- and strong-driving limits: when $A/\omega\rightarrow 0$ we obtain $\Omega_R = A$, and when $A/\omega\rightarrow\infty$ we obtain $\Omega_R=\omega \left| 1 - J_0\left(\frac{2A}{\omega}\right) \right|$, which upon shifting by $\omega$ gives $\Omega_R=-J_0\left(\frac{2A}{\omega}\right)$ (and the minus sign here is physically insignificant because it is the absolute value of this expression that gives the meaningful frequency).

It is worth noting that Ref.~\onlinecite{Lue2012} has an expression for the Rabi frequency that resembles Eq.~(\ref{eq:RabiFrequencyAnalyticalFormulaResonantCase}). The approach used there, however, is designed to extend the regime of validity from the weak-coupling limit to somewhat large driving strengths, and it breaks down in the strong-driving limit. In contrast, our results are most accurate in the weak- and strong-driving limits, with some deviation from the exact results at intermediate values of $A$. Furthermore, the approach of Ref.~\onlinecite{Lue2012} relies on a numerical evaluation of one of the parameters in the argument of the Bessel functions, while our approach is purely analytical.

\begin{figure}[]
\includegraphics[width=80mm]{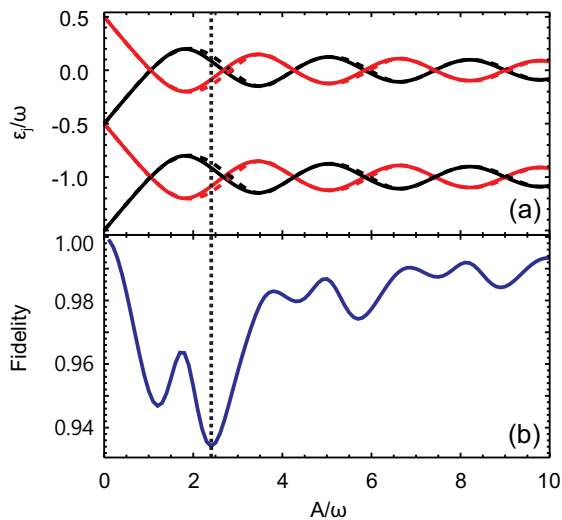}
\caption{(a) The quasienergies as functions of driving amplitude $A$. The black and red lines are quasienergies which correspond to the two inequivalent quasienergies, $\epsilon_{0}$ and $\epsilon_{1}$ respectively. The solid lines are obtained from numerically solving the \sch equation and is essentially exact. The dashed lines are the approximate analytical expressions given by Eq.~\eq{QuasienergiesAnalytical}. (b) The fidelity of the analytical expression for the Floquet modes given by Eq.~\eq{PeriodicPartsOfFloquetStatesOriginalBasis} with respect to the exact Floquet states, which are evaluated by a numerical integration of the \sch equation, as a function of $A$. Away from the first two zeros of $J_0(2A/\omega)$, and in particular for both the weak and the strong driving limits, the analytical formulae give good approximations for the quasienergies and the Floquet modes. The vertical dotted line in (b) marks the point of maximum deviation between the analytical and the exact Floquet modes.}
\label{fig:ComparisonBetweenAnalyticalFormulaAndExactResults}
\end{figure}

In Fig.~\ref{fig:ComparisonBetweenAnalyticalFormulaAndExactResults}, we plot the quasienergies obtained from Eq.~\eq{QuasienergiesAnalytical} and compare them with those obtained from diagonalizing a $100\times 100$ truncated version of the Hamiltonian in Eq.~\eq{FloquetHamiltonian}. We also plot the fidelity $F$ of the Floquet modes obtained from Eq.~\eq{PeriodicPartsOfFloquetStatesOriginalBasis} with the exact Floquet modes:
\begin{equation}
F = \left| \langle\langle u_{0,\rm analytical}(t) | u_{0,\rm exact}(t) \rangle\rangle \right|^2,
\label{eq:FloquetModeFidelity}
\end{equation}
where we have used the definition
\begin{equation}
\langle\langle \cdot \rangle\rangle = \frac{1}{T} \int_{0}^{T} dt \langle \cdot \rangle.
\label{eq:TimeIntegratedInnerProduct}
\end{equation}
The analytical formula for the quasienergies agrees very well with the numerical (and essentially exact) results, except for small errors around the locations of the Floquet-state degeneracy points. The Floquet modes agree reasonably well with the exact results for most values of the driving strength. It should be noted here that a fidelity of 0.93 is not very high when dealing with qubit states: in the time-independent case, this value of the fidelity corresponds to an angle of 30 degrees in the Bloch sphere representation. This difference in the quality of the approximation for the quasienergies and Floquet modes can be understood by considering (non-degenerate) perturbation theory. Perturbations that appear as off-diagonal matrix elements in the Hamiltonian modify the quantum states at the first order but modify the energies only at the second order. In our case these perturbations are the matrix elements that we ignored when we truncated the Floquet Hamiltonian to a $4\times 4$ matrix. For example, taking into consideration the fact that $|J_1(2A/\omega)|$ has maxima of magnitude $\sim 0.5$, we can estimate that an off-diagonal perturbation of magnitude $\delta H \sim\omega J_1(2A/\omega)/2$ coupling a quantum state to another one that is $\sim\omega$ in energy away should give a fidelity between the perturbed and unperturbed states of about $1-\delta F$ where $\delta F\sim\left(\delta H/\omega\right)^2\sim 0.06$, which agrees with the values plotted in Fig.~\ref{fig:ComparisonBetweenAnalyticalFormulaAndExactResults}.

As one can see in Fig.~\ref{fig:ComparisonBetweenAnalyticalFormulaAndExactResults}, there are points where the quasienergies are degenerate, including the point $A=0$. Any superposition of two Floquet states that have the same quasienergy is also a Floquet state, meaning that at these degeneracy points there is ambiguity (and freedom) in how to define the Floquet modes. Since in this paper we are interested in the dynamics under the influence of short pulses with a varying amplitude $A$, the natural definition of the Floquet modes is obtained by considering the path that the pulse takes in parameter space approaching or moving away from any of the degeneracy points. In particular, for a resonant pulse (i.e.~with $\omega=\Delta$) starting from $A=0$ the Floquet states at the initial time should be defined by taking the limit $A\rightarrow 0$. The degeneracy of Floquet states at finite values of $A$ is a result of the symmetry $\cos(\omega(t+T/2))=-\cos(\omega t)$ in the sinusoidal driving waveform that we consider here~\cite{Creffield_2003_FloquetCrossings}. If we modify the waveform and the symmetry is lifted, the quasienergy crossings will turn into avoided crossings.

Before concluding this section, we address a rather surprising aspect in our analysis. Obtaining the expression for the Rabi frequency that is valid in the weak-driving limit was more difficult than obtaining the expression that is valid in the strong-driving limit, although the former limit is generally considered to be the simpler of the two limits. The reason is that we followed a derivation that is well suited for the strong-driving limit and modified it in order to extend its validity to the weak-driving limit. If we were interested in the weak-driving limit only, we could have started the derivation differently and written $H_F$ in Eq.~\eq{FloquetHamiltonian} in the basis $\{\dots , u_{j,n-1,+} , u_{j,n-1,-} , u_{j,n,+} , u_{j,n,-} , u_{j,n+1,+} , \dots \}$ with $\ket{+}=\left(\ket{0}+\ket{1}\right)/\sqrt{2}$ and $\ket{-}=\left(\ket{0}-\ket{1}\right)/\sqrt{2}$:
\begin{widetext}
\begin{equation}
H_F = \left(
\begin{array}{cccccccc}
\ddots & & & & & & & \\
& {(n-1) \omega - \frac{\Delta}{2}} & 0 & 0 & {-\frac{A}{2}} & 0 & 0 & \\
& 0 & {(n-1) \omega + \frac{\Delta}{2}} & {-\frac{A}{2}} & 0 & 0 & 0& \\
& 0 & {-\frac{A}{2}} & {n \omega - \frac{\Delta}{2}} & 0 & 0 & {-\frac{A}{2}} & \\
& {-\frac{A}{2}} & 0 & 0 & {n \omega + \frac{\Delta}{2}} & {-\frac{A}{2}} & 0 & \\
& 0 & 0 & 0 & {-\frac{A}{2}} & {(n+1) \omega - \frac{\Delta}{2}} & 0 & \\
& 0 & 0 & {-\frac{A}{2}} & 0 & 0 & {(n+1) \omega + \frac{\Delta}{2}} & \\
& & & & & & & \ddots
\end{array}
\right).
\label{eq:FloquetHamiltonianInSigmaXBasis}
\end{equation}
\end{widetext}
Near resonance (i.e.~when $\omega=\Delta+\delta$ with $\delta\ll\Delta$) one can obtain a good approximation of the quasienergies and Floquet modes by truncating $H_F$ to the $ 2\times 2$ matrix
\begin{equation}
H_F = \left(
\begin{array}{cc}
{\frac{\Delta}{2}} & {-\frac{A}{2}} \\
{-\frac{A}{2}} & {\frac{\Delta}{2} + \delta} \\
\end{array}
\right).
\end{equation}
which gives the well known expression for the Rabi frequency $\Omega_R=\sqrt{A^2+\delta^2}$. This derivation is indeed simpler than the one that we have followed above. It is not obvious, however, how one could modify this derivation and extend its validity to the strong-driving limit.

\section{\sch Equation and nonadiabatic transitions in the Floquet picture} \label{sec:adiabatic_theory}

We start this section by giving a derivation of the Schr\"odinger equation in the Floquet basis. Before any manipulation, the equation reads
\begin{equation}
i \frac{d}{dt} \ket{\psi(t)} = H \ket{\psi(t)}.
\label{eq:SchroedingerEquation}
\end{equation}
Inspired by the fact that when $A$ is independent of time the probability amplitudes of the Floquet states are also time independent, we express the quantum state $\ket{\psi(t)}$ as a superposition in the instantaneous Floquet basis:
\begin{equation}
\ket{\psi(t)} = \sum_j c_j(t) \ket{\psi_{F,j}(A,t)}, \label{eq:evolution}
\end{equation}
where $\ket{\psi_{F,j}(A,t)}$ is the Floquet state with index $j$ taken at time $t$ and $A$ is assumed to implicitly depend on time. Differentiating this expression with respect to time, we find an alternative expression for the derivative on the left-hand side of Eq.~(\ref{eq:SchroedingerEquation}):
\begin{widetext}
\begin{equation}
\frac{d}{dt} \ket{\psi(t)} = \sum_j \left\{ \frac{dc_j(t)}{dt} \ket{\psi_{F,j}(A,t)} + c_j(t) \frac{dA}{dt} \frac{\partial}{\partial A} \ket{\psi_{F,j}(A,t)} + c_j(t) \frac{\partial}{\partial t} \ket{\psi_{F,j}(A,t)} \right\}.
\label{eq:ExpansionOfPsiDot}
\end{equation}
\end{widetext}
We now note that the partial derivative in the last term corresponds to the full derivative of the Floquet states with respect to time assuming that $A$ remains constant in time. With this point in mind, we know that the Floquet states are solutions of the \sch equation:
\begin{equation}
i \frac{\partial}{\partial t} \ket{\psi_{F,j}(A,t)} = H \ket{\psi_{F,j}(A,t)}.
\end{equation}
The last term in Eq.~(\ref{eq:ExpansionOfPsiDot}) therefore cancels the right-hand side in Eq.~(\ref{eq:SchroedingerEquation}), and the \sch equation in the Floquet basis reduces to
\begin{equation}
\sum_j \left\{ \frac{dc_j(t)}{dt} \ket{\psi_{F,j}(A,t)} + c_j(t) \frac{dA}{dt} \frac{\partial}{\partial A} \ket{\psi_{F,j}(A,t)} \right\} = 0.
\end{equation}
Multiplying this equation on the left by $\bra{\psi_{F,k}(A,t)}$, we find that
\begin{equation}
\frac{dc_k(t)}{dt} = - \sum_j c_j(t) \frac{dA}{dt} \bra{\psi_{F,k}(A,t)} \frac{\partial}{\partial A} \ket{\psi_{F,j}(A,t)},
\end{equation}
which can alternatively be expressed as
\begin{widetext}
\begin{eqnarray}
\frac{dc_k(t)}{dt} & = & - \sum_j c_j(t) \frac{dA}{dt} \left( -it \frac{d\epsilon_j}{d A} \delta_{kj} + e^{i(\epsilon_k-\epsilon_j)t} \bra{u_{k}(A,t)} \frac{\partial}{\partial A} \ket{u_{j}(A,t)} \right) \nonumber \\
& = & it \frac{d\epsilon_k}{dt} c_k(t) - \frac{dA}{dt} \sum_j c_j(t) e^{i(\epsilon_k-\epsilon_j)t} \bra{u_{k}(A,t)} \frac{\partial}{\partial A} \ket{u_{j}(A,t)}.
\label{eq:SchroedingerEquationInFloquetBasis}
\end{eqnarray}
\end{widetext}
The two terms in the above equation have clear meanings. The first term describes the fact that as $A$ changes (and $\epsilon$ changes with it) the coefficients $c_k$ acquire phase shifts in order to correct for the difference between the actually accumulated phase (from the initial time until time $t$) and the phase that would have accumulated assuming that the instantaneous quasienergy had been in effect from $t=0$ until time $t$. The second term on the right-hand side of Eq.~(\ref{eq:SchroedingerEquationInFloquetBasis}) describes geometric phase accumulation, as well as nonadiabatic transitions between Floquet states.

The first term in Eq.~\eq{SchroedingerEquationInFloquetBasis} has the undesirable property that its coefficient contains the factor $t$, which grows without bounds, a feature that looks rather unnatural and could complicate calculations for long pulses. This term can be eliminated by defining the coefficients
\begin{equation}
\tilde{c}_k(t)= e^{-i[t\epsilon_k(t)-\int_0^t\epsilon_k(t')dt']} c_k(t). \label{eq:c2ctilde}
\end{equation}
In terms of the coefficients $\tilde c_j(t)$, the quantum state in Eq.~\eq{evolution} takes the form
\be
\ket{\psi(t)} = \sum_j \tilde c_j(t) \ket{u_j(A,t)}e^{-i\int_0^t\epsilon_j(t')dt'}. \label{eq:evolution_tildec}
\ee
This equation exhibits a clear analogy with quantum state evolution in standard adiabatic theory: the Floquet modes $\ket{u_j(A,t)}$ are the eigenstates of the instantaneous Floquet Hamiltonian, the dynamical phases $\int_0^t\epsilon_j(t')dt'$ are the time integrals of the quasienergies, and the evolution of the coefficients $\tilde c_j(t)$ incorporates geometric phases and nonadiabatic transitions.

With the definition of $\tilde c_j(t)$, Eq.~\eq{SchroedingerEquationInFloquetBasis} is transformed into
\begin{widetext}
\be
\frac{d\tilde{c}_k(t)}{dt} = - \frac{dA}{dt} \sum_j e^{-i\int_0^t[\epsilon_j(t')-\epsilon_k(t')]dt'} \bra{u_{k}(A,t)} \frac{\partial}{\partial A} \ket{u_{j}(A,t)} \tilde{c}_j(t). \label{eq:SchroedingerEquationInFloquetBasisWithTildes}
\ee
\end{widetext}
Note that although we have eliminated the indefinitely growing coefficient with this last transformation we now have a new undesirable feature in the equation of motion, namely the fact that the time derivative of the state now depends on the entire history of the system, as opposed to just the system parameters and state at time $t$.

Eq.~\eq{SchroedingerEquationInFloquetBasisWithTildes} is the \sch equation that describes the transition dynamics between the Floquet modes. If the driving strength $A$ changes very slowly over time $t$, we are in the adiabatic limit of the Floquet picture. Using Eq.~\eq{SchroedingerEquationInFloquetBasisWithTildes} together with $d A/d t \rightarrow 0$, we conclude that all the coefficient $\tilde c_{k}(t)$ are constants. In this case, there is no population transition between the Floquet modes. From Eq.~\eq{evolution_tildec}, we find that the state dynamics are fully described by applying a time-dependent dynamical phase $-\int_0^t\epsilon_j(t')dt'$ to each of the two Floquet modes $\ket{u_j(A,t)}$. The constant coefficients $\tilde c_j$ can be obtained by decomposing the initial state $\ket{\psi(0)}$ in the Floquet state basis at the initial time, e.g.~the basis states $\ket{u_j(0,0)}$ if we are considering a pulse whose amplitude $A$ starts from zero at the initial time $t=0$ (and we remind the reader here that setting $A=0$ in $\ket{u_j(0,0)}$ should be understood as taking the limit $A\rightarrow 0$).

\begin{figure}[]
\includegraphics[width=80mm]{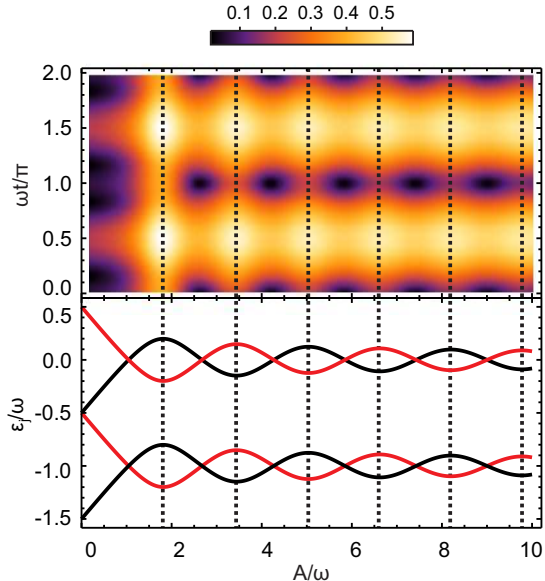}
\caption{The dimensionless  quantity $\omega|\bra{u_{0}(A,t)} \frac{\partial}{\partial A} \ket{u_{1}(A,t)}|$ (upper panel) and the quasienergy spectrum (lower panel) for the case of a qubit under resonant driving, $\omega = \Delta$. The quasienergies and Floquet modes are calculated by numerically diagonalizing the Floquet Hamiltonian, Eq.~\eq{FloquetHamiltonian}. The positions of the avoided crossings (labeled with dashed lines in the lower panel) align well with the local maxima of $|\bra{u_{0}(A,t)} \frac{\partial}{\partial A} \ket{u_{1}(A,t)}|$.}
\label{fig:InnerProduct}
\end{figure}

If the driving strength $A$ varies fast over time, nonadiabatic transitions between the Floquet modes can be excited. According to Eq.~\eq{SchroedingerEquationInFloquetBasisWithTildes}, the nonadiabatic transition rate, given by the change of the coefficient $d\tilde c_k/dt$, depends on the inner product $\bra{u_{k}(A,t)} \frac{\partial}{\partial A} \ket{u_{j}(A,t)}$. In the upper panel of Fig.~\ref{fig:InnerProduct}, we plot the amplitude of the inner product $|\bra{u_{0}(A,t)} \frac{\partial}{\partial A} \ket{u_{1}(A,t)}|$ for the case of a qubit under resonant driving, $\omega = \Delta$. This quantity exhibits multiple local maxima at driving amplitudes which correspond to the positions of the avoided crossings of the quasienergies spectrum (labeled with dashed lines in the lower panel of Fig.~\ref{fig:InnerProduct}). In a pulse where the driving strength changes and the system traverses such an avoided crossing, Landau-Zener transitions can occur between Floquet states following the usual formula for Landau-Zener transitions, just modified from the language of energies to that of quasienergies.

\subsection*{An alternative form of the \sch equation}

The time dependence of the Floquet modes $\ket{u_{j}(A,t)}$ appears explicitly in Eq.~\eq{SchroedingerEquationInFloquetBasisWithTildes}. It is possible to rewrite the equation using the Fourier decomposition of these modes, similarly to what is done for example when transforming the time-dependent Floquet Hamiltonian into its time-independent infinite matrix form. As we shall see shortly, this change makes it possible to evaluate the necessary inner products directly from the eigenstates of the Floquet Hamiltonian in its time-independent large-matrix form. In order to perform this transformation, we take the inner product on the right-hand side of Eq.~\eq{SchroedingerEquationInFloquetBasisWithTildes} and express it as a Fourier series:
\begin{widetext}
\begin{eqnarray}
\bra{u_{k}(A,t)} \frac{\partial}{\partial A} \ket{u_{j}(A,t)} & = & \sum_{n=-\infty}^{\infty} e^{in\omega t} \frac{1}{T} \int_{0}^{T} dt' e^{-in\omega t'} \bra{u_{k}(A,t')} \frac{\partial}{\partial A} \ket{u_{j}(A,t')} \nonumber \\
& = & \sum_{n=-\infty}^{\infty} e^{in\omega t} \langle\langle u_{k}(A,t') | \frac{\partial}{\partial A} | u_{j}^{(n)}(A,t') \rangle\rangle
\end{eqnarray}
\end{widetext}
where the double-bracket notation is defined in Eq.~\eq{TimeIntegratedInnerProduct}, and we have also used the definition
\begin{equation}
\ket{u_{j}^{(n)}(A,t)} = e^{-in\omega t} \ket{u_{j}(A,t)}.
\end{equation}
Note that $\ket{u_j^{(n)}(A,t)}$ is different from $\ket{u_{j,n}(A)}$, which was introduced in Sec.~\ref{sec:Floquet} and is the $n$-th Fourier coefficient of $\ket{u_j(A,t)}$. Apart from the differentiation with respect to $A$, the inner products in Eq.~\eq{SchroedingerEquationInFloquetBasisWithTildes} can be evaluated by taking the simple inner products between two vectors that are obtained by diagonalizing the Hamiltonian $H_F$. Because of the ambiguity in defining $\epsilon_j$ and $\ket{u_j(t)}$ (more specifically the fact that one can freely move integer multiples of $\omega$ between the two), the above definition of $\ket{u_{j}^{(n)}(A,t)}$ naturally suggests the accompanying definition
\begin{equation}
\epsilon_{j}^{(n)} = \epsilon_j - n\omega,
\end{equation}
which leads to the relation
\begin{equation}
e^{-i\epsilon_{j}^{(n)}t} \ket{u_{j}^{(n)}(A,t)} = e^{-i\epsilon_{j}t} \ket{u_{j}(A,t)},
\end{equation}
for all values of $n$. In other words, the different values of $n$ give identical copies of the Floquet states, although the quasienergies and Floquet modes for the different values of $n$ are different by integer multiples of $\omega$ and a factor of $e^{-in\omega t}$. Equation \eq{SchroedingerEquationInFloquetBasisWithTildes} now becomes
\begin{widetext}
\begin{eqnarray}
\frac{d\tilde{c}_k(t)}{dt} & = & - \frac{dA}{dt} \sum_j \sum_{n=-\infty}^{\infty} e^{-i\int_0^t[\epsilon_j(t')-\epsilon_k(t')]dt'} e^{in\omega t} \langle\langle  u_{k}(A,t) | \frac{\partial}{\partial A} | u_{j}^{(n)}(A,t) \rangle\rangle \tilde{c}_j(t) \nonumber \\
& = & - \frac{dA}{dt} \sum_j \sum_{n=-\infty}^{\infty} e^{-i\int_0^t[\epsilon_{j}^{(n)}(t')-\epsilon_k(t')]dt'} \langle\langle  u_{k}(A,t) | \frac{\partial}{\partial A} | u_{j}^{(n)}(A,t) \rangle\rangle \tilde{c}_j(t).
\end{eqnarray}
This equation is asymmetric in that it uses the quasienergies and Floquet modes both with and without the additional index $n$. It can be made symmetric by first rewriting it in the form
\begin{eqnarray}
\frac{d\tilde{c}_{k}^{(0)}(t)}{dt} & = & - \frac{dA}{dt} \sum_j \sum_{n=-\infty}^{\infty} e^{-i\int_0^t[\epsilon_{j}^{(n)}(t')-\epsilon_{k}^{(0)}(t')]dt'} \langle\langle  u_{k}^{(0)}(A,t) | \frac{\partial}{\partial A} | u_{j}^{(n)}(A,t) \rangle\rangle \tilde{c}_{j}^{(0)}(t)
\label{eq:SchroedingerEquationInFloquetBasisWithTildesAndSpaceExpansionUnsymmetrized}
\end{eqnarray}
\end{widetext}
and observing that the equation would still be valid if we replaced the index 0 by $m$ throughout the equation. We can therefore replace the single equation by an infinite number of equivalent and independent equations. Because these equations are linear in the coefficients $\tilde{c}$, one can distribute the two probability amplitudes $\tilde{c}_j$ at the initial time (with complete freedom) among the different identical copies of the Floquet states (and hence use the coefficients $\tilde{c}_j^{(n)}$ with $n=-\infty,\dots,\infty$) and solve all the equations in order to find the probability amplitudes at the final time, keeping in mind that after obtaining the probability amplitudes at the final time all the coefficients with the same value of $j$ must be summed before calculating the occupation probabilities for the two Floquet states, i.e. $\tilde{c}_j(t) = \sum_{n=-\infty}^{\infty} \tilde{c}_j^{(n)}(t)$. Noting once more the linearity in the \sch equations and the equivalence between the different copies of Floquet state, we can rearrange the terms on the right-hand side among the different equations. One possible rearrangement gives the set of equations
\begin{widetext}
\begin{eqnarray}
\frac{d\tilde{c}_{k}^{(m)}(t)}{dt} & = & - \frac{dA}{dt} \sum_j \sum_{n=-\infty}^{\infty} e^{-i\int_0^t[\epsilon_{j}^{(n)}(t')-\epsilon_{k}^{(m)}(t')]dt'} \langle\langle  u_{k}^{(m)}(A,t) | \frac{\partial}{\partial A} | u_{j}^{(n)}(A,t) \rangle\rangle \tilde{c}_{j}^{(n)}(t).
\label{eq:SchroedingerEquationInFloquetBasisWithTildesAndSpaceExpansion}
\end{eqnarray}
\end{widetext}
We remark that the above equation can also be derived using the ($t, t'$) formalism~\cite{Peskin} as described in Ref.~\onlinecite{Drese:1999tq}.

\section{Qubit final state after pulses with slowly varying amplitude} \label{sec:adiabatic}
In applications to quantum control of a single qubit, one often focuses on the qubit's final state at the end of a control pulse or pulse sequence, which is then followed by state readout. In this section, we discuss the final state of the qubit after a resonant pulse whose amplitude varies slowly. In this case, the adiabatic limit in the Floquet picture applies. The pulse amplitude $A=0$ at the beginning and the end of the pulse. As a result, we would like to express the initial and final states of the qubit using Floquet modes $\ket{u_j(0,t)}$. From Eqs.~\eq{BasicFormOfFloquetStates}, \eq{theta} and \eq{PeriodicPartsOfFloquetStatesOriginalBasis}, we obtain the Floquet modes expressions
\ba
\ket{u_0(0,t)} &= \frac{1}{\sqrt{2}}\left(
                   \begin{array}{c}
                     1 \\
                     e^{i\omega t} \\
                   \end{array}
                 \right), \nn \\
\ket{u_1(0,t)} &= \frac{1}{\sqrt{2}}\left(
                   \begin{array}{c}
                     1 \\
                     -e^{i\omega t} \\
                   \end{array}
                 \right)
\ea
in the original qubit basis using the Bessel functions at value 0, $J_n(0) = \delta_{0n}$, and $\theta = \pi/2$ for $\omega = \Delta$ and $A\rightarrow 0$. We notice from the above equations that the Floquet modes $\ket{u_j(0,t)}$ are the time-independent states $\frac{1}{\sqrt{2}}(\ket{\tilde 0}+\ket{\tilde 1})$ and $\frac{1}{\sqrt{2}}(\ket{\tilde 0}-\ket{\tilde 1})$, if we define $\ket{\tilde 0}$ and $\ket{\tilde 1}$ as the qubit eigenstates in a rotating frame which rotates at an angular frequency $\omega$. In this rotating frame, the Floquet modes at the beginning and at the end of the pulse correspond to the same set of states. The state evolution from the beginning ($t=0$) to the end ($t=t_f$) of the pulse is simply described by the accumulation of a dynamical phase $\phi_d =\int_0^{t_f}\Delta \epsilon(t')dt'$, where $\Delta\epsilon(t) = \epsilon_2(t) - \epsilon_1(t)$. In the Bloch sphere, the above evolution can be seen as a rotation of the state vector by an angle $\phi_d$ around the x axis, which points towards $\ket{u_0(0,t)}$. The rotation angle $\phi_d$, which is the time integral of the quasienergy difference, does not depend on any other details of the pulse shape. Considering the case where the rise and fall parts of the pulse are slow and we vary only the duration of the middle part of the pulse with fixed amplitude $A_m$, the qubit state undergoes Rabi oscillations as $\phi_d$ changes at a constant rate $\Omega_R = \Delta \epsilon(A_m)$, i.e.~the Rabi frequency as discussed in Sec.~\ref{sec:Floquet}.

\begin{figure}[]
\includegraphics[width=80mm]{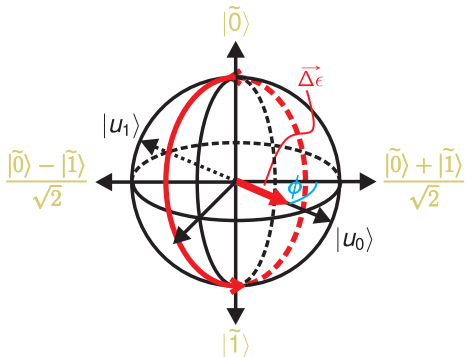}
\caption{Bloch sphere representation of the adiabatic evolution in the Floquet picture. The state vector evolution (red thin arrows) is a rotation around a fictitious field (red thick arrows) pointing towards the direction of $\ket{u_0(0,t)}$ in the equatorial plane. The direction of the fictitious field in the equatorial plane is determined by the phase $\phi$ of the driving and the amplitude of the fictitious field has an amplitude given by the quasienergy difference $\Delta\epsilon$.}
\label{fig:AdaibaticPulseBlochRepresentation}
\end{figure}
It is worth noting here that although we have only discussed the case where the driving term of the qubit is $A(t)\cos(\omega t)$ above, we can easily generalize our results to the driving waveform $A(t)\cos(\omega t + \phi)$ with an arbitrary constant phase $\phi$. In this more general case, we obtain the same form of the quasienergies of the system and Floquet modes with a phase shift compared to the $\phi=0$ case. In particular, the Floquet modes at $A\rightarrow 0$ are given by $\ket{u_{0}(0,t)} = \frac{1}{\sqrt{2}}(\ket{\tilde 0}+e^{i\phi}\ket{\tilde 1})$ and $\ket{u_{1}(0,t)} = \frac{1}{\sqrt{2}}(\ket{\tilde 0}-e^{i\phi}\ket{\tilde 1})$. By choosing a specific value of $\phi$ in the driving signal, one can control the rotation axis of the Rabi oscillation in the equatorial plane of the Bloch sphere (see Figure~\ref{fig:AdaibaticPulseBlochRepresentation}). This point appears naturally in the RWA for weak driving, and our analysis shows that it holds for pulses with arbitrary driving amplitude, as long as the pulse amplitude varies slowly. In Sec.~\ref{sec:nonadiabatic}, we will examine more closely the breakdown of the adiabatic approximation.

\section{Nonadiabatic transitions between Floquet states of a qubit} \label{sec:nonadiabatic}

\subsection*{Dynamics with adiabatic perturbation theory}

When the driving strength changes rapidly, e.g.~in a short pulse, nonadiabatic transitions between Floquet states will be reflected in the evolution of the coefficients $\tilde c_j(t)$ in Eq.~\eq{evolution_tildec}. This evolution can be calculated using Eq.~\eq{SchroedingerEquationInFloquetBasisWithTildesAndSpaceExpansion} or equivalently Eq.~\eq{SchroedingerEquationInFloquetBasisWithTildes}. These two equations are exact. To proceed further analytically, we will employ adiabatic perturbation theory (APT)~\cite{Rigolin2008} and then we will compare the solutions given by APT with the exact numerical solutions. In the first order of APT~\cite{Drese:1999tq, Rigolin2008}, we solve Eq.~\eq{SchroedingerEquationInFloquetBasisWithTildesAndSpaceExpansion} iteratively keeping terms up to first order in $dA/dt$. This is equivalent to replacing $\tilde c_{j,n}(t) \rightarrow \tilde c_{j,n}(0)$ on the right-hand side of Eq.~\eq{SchroedingerEquationInFloquetBasisWithTildesAndSpaceExpansion}. This way we obtain the approximate expression
\begin{widetext}
\be
\tilde{c}_{k}^{(m)}(t) = \tilde{c}_{k}^{(m)}(0) - \int_0^{t} dt \frac{dA}{dt} \sum_j \sum_{n=-\infty}^{\infty} e^{-i\int_0^t[\epsilon_{j}^{(n)}(t')-\epsilon_{k}^{(m)}(t')]dt'} \langle\langle  u_{k}^{(m)}(A,t) | \frac{\partial}{\partial A} | u_{j}^{(n)}(A,t) \rangle\rangle \tilde{c}_{j}^{(n)}(0).
\label{eq:SchrodingerEquation1stOrderAPT}
\ee
Setting $\tilde{c}_{j}^{(n)}(0)=\tilde{c}_{j}(0) \delta_{n,0}$ at the initial time and summing the equation over $m$, we obtain
\be
\tilde{c}_{k}(t) = \tilde{c}_{k}(0) - \int_0^{t} dt \frac{dA}{dt} \sum_j \sum_{m=-\infty}^{\infty} e^{-i\int_0^t[\epsilon_{j}^{(0)}(t')-\epsilon_{k}^{(m)}(t')]dt'} \langle\langle  u_{k}^{(m)}(A,t) | \frac{\partial}{\partial A} | u_{j}^{(0)}(A,t) \rangle\rangle \tilde{c}_{j}(0).
\label{eq:1stOrderAPT}
\ee
If we quantify the nonadiabatic transition as the change of each coefficient $\delta \tilde c_j(t) = \tilde c_j(t) - \tilde c_j(0)$, we can see from the above equation that this quantity is proportional to
\be
N_{j\rightarrow k}(t) = \int_0^{t} dt \frac{dA}{dt} \sum_{m=-\infty}^{\infty} e^{-i\int_0^t[\epsilon_{j}^{(0)}(t')-\epsilon_{k}^{(m)}(t')]dt'} \langle\langle  u_{k}^{(m)}(A,t) | \frac{\partial}{\partial A} | u_{j}^{(0)}(A,t) \rangle\rangle. \label{eq:NonadAmp}
\ee
The above quantity, $N_{j\rightarrow k}(t)$, denotes the nonadiabatic transition matrix element from the $j$-th to the $k$-th Floquet mode. It can be defined as the complex coefficient $\tilde c_k$ of the $k$-th Floquet mode after a driven evolution with the initial state prepared in the $j$-th Floquet mode, i.e. $\tilde c_l(0) = \delta_{lj}$. $N_{j\rightarrow k}$ is the sum of complex, oscillating terms that will occasionally give a very small value for the sum, as we shall see below. We also note that the coefficient $\tilde c_k$ can be calculated essentially exactly by numerically solving the time-dependent \sch equation.

Next, we calculate the nonadiabatic transition matrix elements for a representative pulse shape that has the simple form:
\be
A(t) = \begin{cases}
\frac{A_{m}}{2}(1-\cos(\pi t/t_{r})) & t \le t_{r} \\
A_{m} & t_{r} < t \le t_{r} + t_{p} \\
\frac{A_{m}}{2}(1+\cos(\frac{\pi (t-t_{p}-t_{r})}{t_{f}})) & t_{r} + t_{p} < t \le t_{r}+t_{p}+t_{f}
\end{cases}, \label{eq:pulseshape}
\ee
\end{widetext}
where $t_{p}$ is the duration of the middle part of the pulse, $t_{r}$ ($t_f$) is the rise (fall) time of the pulse, and $A_{m}$ is the maximum amplitude of the pulse. We choose to analyze pulses with the rising and falling edge being cosine functions because it is a pulse shape that requires a small bandwidth, minimizing required experimental resources. Nonadiabatic transitions occur in both the rising and falling edges of the pulse. We will deal with these two cases separately in the remainder of this section. We will consider the case of resonant driving ($\omega=\Delta$) and $\phi=0$ here but our treatment can be generalized to arbitrary values of $\omega$ and $\phi$.

\subsection*{Rising edge}
\begin{figure*}[]
  \includegraphics[width=180mm]{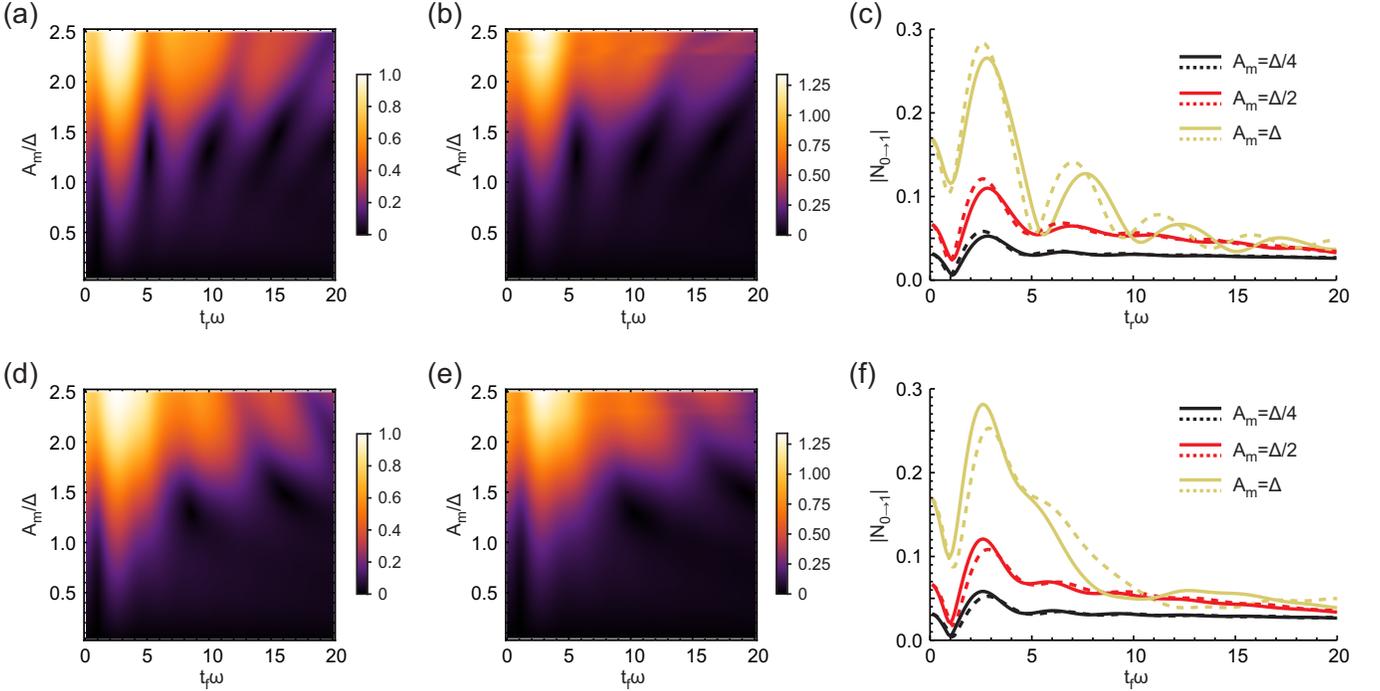}
  \caption{Nonadiabatic transition amplitude between the two Floquet modes of a qubit induced by ramping the pulse amplitude on the rising (a-c) and falling (d-f) edges of the pulse. The qubit is driven on resonance ($\omega = \Delta$) with $\phi = 0$ and pulse shapes given by Eq.~\eq{pulseshape}. (a,d) The nonadiabatic transition amplitude calculated by setting $\tilde c_0(0) = 1$ and numerically solving the time-dependent \sch equation for $\tilde c_1(t)$ plotted as a function the maximum driving amplitude $A_{m}$ and the pulse rise/fall time $t_{r/f}$. (b,e) The nonadiabatic transition amplitude $|N_{0\rightarrow 1}|$ calculated using adiabatic perturbation theory (APT), Eq.~\eq{NonadAmp}. (c,f) Comparison between the exact solutions (solid) and the APT solutions (dashed) at $A_m = \Delta/4$, $\Delta/2$, and $\Delta$. All the curves in panels c and f have a clear minimum in the nonadiabatic transition amplitude at $\omega t_{r/f}\approx 1$.} \label{fig:nonadiabatic}
\end{figure*}

For the pulse shape given by Eq.~\eq{pulseshape}, the rising edge is in the time interval between 0 and $t_r$. For performing the exact numerics to simulate the nonadiabatic transitions, we assume the initial state of the system to be $\ket{\psi(0)} = (\ket{0} + \ket{1})/\sqrt{2}$ which coincides with the Floquet mode $\ket{u_0(0,0)}$. Thus $\tilde c_0(0) = 1$ and $\tilde c_1(0) = 0$. We then use a numerical solver to determine the system evolution under the time-dependent Hamiltonian in Eq.~\eq{Hamstart} with the specified pulse shape. The final state $\ket{\psi(t_r)}$ is written as a superposition of the Floquet modes $\ket{u_j(A_m, t_r)}$, yielding the Floquet mode coefficients $\tilde c_0(t_r)$ and $\tilde c_1(t_r)$. In this case, the quantity $\tilde c_1(t_r)$ gives an measure of the nonadiabatic transitions during the rising edge of the pulse. In Fig.~\ref{fig:nonadiabatic}(a), we show $\tilde c_1(t_r)$ as a function of the maximum amplitude $A_{m}$ and the rise time $t_{r}$. For comparison we also evaluate the nonadiabatic transition amplitudes using APT. For this purpose, we first numerically calculate all the instantaneous quasienergies and Floquet modes for values of $A$ from $t=0$ to $t=t_r$ in small steps. We then numerically evaluate $N_{0\rightarrow 1}$ according to Eq.~\eq{NonadAmp}, and we plot its absolute value in Fig.~\ref{fig:nonadiabatic}(b). In Fig.~\ref{fig:nonadiabatic}(c), we show the nonadiabatic transition amplitude calculated both by exact numerics and by APT as a function of $t_{r}$ for a few values of the strong-driving pulse amplitude $A_{m}$. We find good agreement between the APT solution and the exact results. Considering the physics of Landau-Zener transitions, and noting that the minimum gap in the first avoided crossing of quasienergies is $\sim\omega$, we find that for $A_{m}\sim\Delta$ adiabaticity is determined by the product $\omega t_{r}$. As expected, for long rise times $t_{r} \gg 1/\omega$ where the evolution should be adiabatic, we have $|N_{0\rightarrow 1}| \rightarrow 0$. For short rise times $t_{r} \sim 1/\omega$, the nonadiabatic transition amplitude oscillates as a function of $t_{r}$. This oscillatory behavior is a result of interference between different paths that correspond to the nonadiabatic transitions during the rise time. We notice that for the particular pulse shape analyzed here, $|N_{0\rightarrow 1}|$ has a local minimum around $t_{r} \approx 1/\omega$ (as well as at other higher points), which only weakly depends on the value of $A_{m}$ provided that $A_{m} \lesssim \Delta$. At this point, the net probability for a nonadiabatic transition to have occurred during the amplitude ramp is remarkably small. This suppression of nonadiabatic transitions can be understood as destructive interference of the transitions between the Floquet states at different times during the amplitude ramp. The exact values of $t_r$ at which the minima occur depend on $\phi$. For $\phi=\pi/2$ for example, the first minimum in the nonadiabatic transition probability occurs around $t_r \approx 2.6/\omega$.

\subsection*{Falling edge}

The nonadiabatic transitions in the falling edge of the pulse can be calculated in a similar way as for the rising edge. To simplify notations, we shift the times so that the beginning and end times are $0$ and $t_f$ (e.g.~by taking $t_r = t_p = 0$ in Eq.~\eq{pulseshape}). We have
\be
\frac{dA}{dt} = -\frac{\pi A_{m}}{2t_{f}}\sin\l( \arccos \l(\frac{2A(t)}{A_{m}}-1 \r)\r).
\ee
To exactly simulate the nonadiabatic transitions, we start with the initial state $\ket{\psi(0)} = \ket{u_0(A_m,0)}$ and express the final state $\ket{\psi(t_f)}$ in the basis of Floquet modes $\ket{u_j(0, t_f)} = (\ket{0}\pm e^{i\omega t_f}\ket{1})/\sqrt{2}$.
In Fig.~\ref{fig:nonadiabatic}(d-f) we show the amplitude of nonadiabatic transitions occurring during the falling edge of the pulse as a function of the maximum pulse amplitude $A_{m}$ and the fall time $t_{f}$. In the adiabatic limit $t_{f}\gg 1/\omega$, we have $|N_{0\rightarrow 1}| \rightarrow 0$ as expected. We also notice a coherent suppression of nonadiabatic transitions at $t_{f}\approx 1/\omega$ with only a weak dependence on the value of $A_m$ as long as $A_{m} \lesssim \Delta$.

\subsection*{Efficient suppression of transitions between Floquet states}

Our results for the rising and falling edges agree in predicting that when $t_{r}, t_{f}\gg 1/\omega$, nonadiabatic transitions are negligible, which corresponds to the adiabatic limit in this driven system. The small differences between the two cases in the exact shapes of the oscillatory behavior of $|N_{0\rightarrow 1}| \rightarrow 0$ as a function of $t_{r}$ is not very surprising, given that there is an asymmetry between the two cases, both in the pulse shape and in the fact that in one case the pulse amplitude is minimum at the initial time while in the other the pulse amplitude is maximum at the initial time.

Remarkably, we find that the transitions between the Floquet states during the rise and fall edges of the control pulses are reduced for specific values of the rise and fall times, respectively. This effect provides the basis for a quantum control protocol, which we name Floquet Interference Efficient Suppression of Transitions in the Adiabatic basis (FIESTA). In the next section we discuss the application of FIESTA to optimization of single qubit gates.

\section{Application to single-qubit gates} \label{sec:gates}

As we have discussed in Secs.~\ref{sec:adiabatic} and \ref{sec:nonadiabatic}, the absence of nonadiabatic transitions between Floquet modes results in smooth and simple Rabi oscillations between energy eigenstates of the qubit. This is a desirable property for implementing high-fidelity single-qubit gates. Nonadiabatic transitions can be suppressed either by turning the pulse on and off adiabatically or by using the destructive interference of the nonadiabatic transitions discussed in Sec.~\ref{sec:nonadiabatic}. In any realistic setup, the gate fidelity is further reduced by the decoherence that the qubit experiences during the implementation of the gate, and the reduction increases with increasing gate time. For an adiabatic pulse in the strong driving regime, we require $t_{r}, t_{f} \gg 1/\Delta$, which means that the total pulse duration must satisfy $t_{\t{total}} \gg 1/\Delta$. This condition is the same one obtained for the implementation of a typical single-qubit gate in the weak driving regime, because the pulse duration in that case is given by $t_{\t{total}} A_{m} \sim 1$ and $A_{m} \ll \Delta$. Therefore, adiabatic pulses in the strong driving regime do not necessarily yield higher operation speeds than weak-driving pulses. However, with the coherent suppression of the nonadiabatic transitions, which requires careful control of the pulse shapes, one could realize high-fidelity single-qubit gates with gate times on the order of $t_{\t{total}} A_{m} \sim 1/\Delta$. In this section, we discuss the optimization of the fidelity of single-qubit operations by zeroing the nonadiabatic transitions between Floquet modes in fast pulses.

Universal single-qubit operations require unitary operations corresponding to rotations around two axes in the Bloch sphere with arbitrary angles. Here, we consider implementing operations which are rotations around axes in the equatorial plane of the Bloch sphere since these can be realized with the Hamiltonian in Eq.~\eq{HamRF} in the absence of nonadiabatic transitions between Floquet modes. The rotation axis is controlled by the phase of the pulse $\phi$ and the rotation angle $\theta$ is determined by the time integral of the quasienergy difference, i.e. $\theta = \int_0^tdt' \Delta\epsilon(t')$. Following the discussions in the previous section, we consider the pulse shape defined in Eq.~\eq{pulseshape}. To achieve the above rotations, we choose the optimal rise time $t_r$ and fall time $t_f$ on the rising and falling edges respectively for the coherent suppression of nonadiabatic transitions and we choose the duration $t_p$ of the middle part of the pulse correspondingly according to the target rotation angle $\theta$.

\begin{figure*}[]
  \includegraphics[width=160mm]{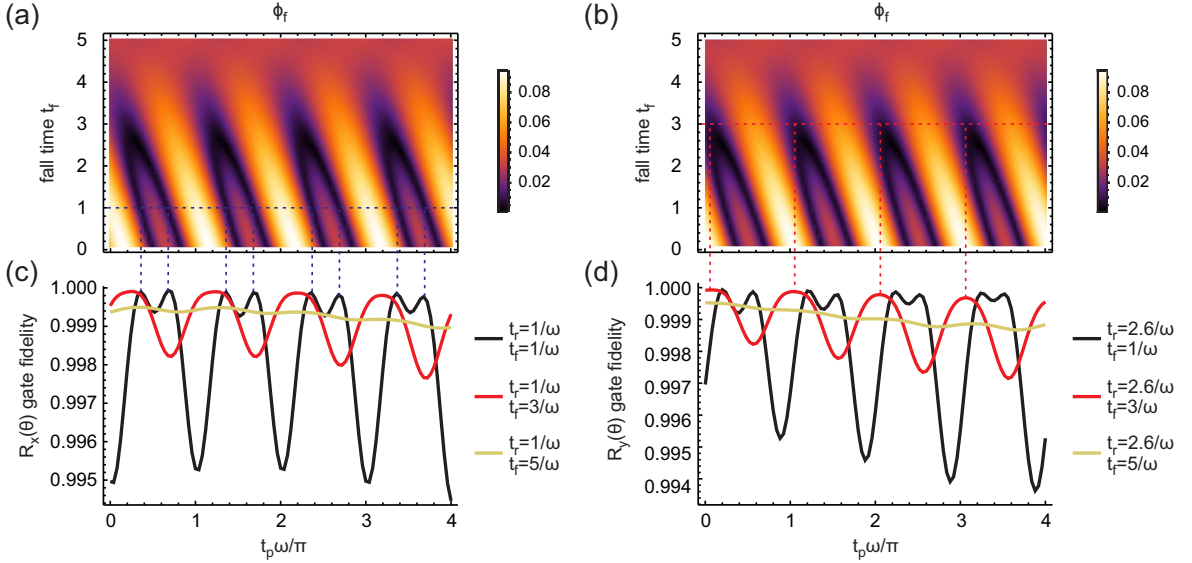}
  \caption{Nonadiabatic transitions between Floquet modes on the falling edge and gate fidelities of the pulses for $R_x(\theta)$ with $\phi=0$ (a,c) and $R_y(\theta)$ with $\phi=-\pi/2$ (b,d). The maximum amplitude of the pulses are assumed to be $A_m = \Delta/4$. The falling edge for the nonadiabatic transition calculation are assumed to be in the time interval between $t_r + t_p$ and $t_r + t_p + t_f$. (a,b) Nonadiabatic transitions on the falling edge of the pulses versus $t_p$ and $t_f$. (c,d) Simulated gate fidelity of $R_x(\theta)$ (c) and $R_y(\theta)$ (d) operations versus $t_p$ with different values of $t_f$. The rise times $t_r$ are chosen to be $1/\omega$ and $2.6/\omega$ for $R_x(\theta)$ and $R_y(\theta)$ respectively in order to achieve optimal suppressions of the nonadiabatic transitions on the rising edge. The optimal pulse parameters that lead to maximum gate fidelities correspond to where the nonadiabatic transitions on the both the rising and falling edge are coherently suppressed.} \label{fig:pulsefidelity}
\end{figure*}
For finding the optimal rise time $t_r$ for a given operation, we calculate and minimize the nonadiabatic transitions on the rising edge in its time interval between 0 and $t_r$ for pulses with a initial phase $\phi$ which sets the desirable rotation axis. This is essentially the same calculation we did in Sec.~\ref{sec:nonadiabatic}. For operations $R_x(\theta)$ and $R_y(\theta)$ which are rotations around the $x$ and $y$ axes of the Bloch sphere, the optimal rise times $t_r$ are approximately $1/\omega$ and $2.6/\omega$ respectively. For finding the optimal fall time $t_f$, the procedure is more complicated as the nonadiabatic transitions on the falling edge depend on $t_f$ as well as the phase of the pulse at the beginning of the falling edge $\phi_f$. This phase depends on the initial phase of the pulse as well as the pulse duration by $\phi_f = \phi + \omega (t_r + t_p)$. In Fig.~\ref{fig:pulsefidelity}(a) and (b), we plot the nonadiabatic transition amplitude $|N_{0\rightarrow 1}|$ on the falling edge versus the fall time $t_f$ and the phase $\phi_f$ with the maximum amplitude $A_m = \Delta/4$. We find that the $t_f$ minimizing $|N_{0\rightarrow 1}|$ changes with $\phi_f$ periodically.

Next, we perform simulation on the qubit dynamics under pulses with different parameters and compute the their fidelities with respect to the ideal operations $R_x(\theta)$ and $R_y(\theta)$. With the shape of the pulse in Eq.~\eq{pulseshape}, the pulse is fully determined by the parameters $A_{m}$, $\phi$, $t_{r}$, $t_{f}$, and $t_{p}$. With energy relaxation time $T_1$ taken into account, we can express the resulting operation as a quantum process matrix $\chi(A_{m}, \phi, t_{r}, t_{f}, t_{p}, T_1)$ characterized by quantum process tomography (QPT)~\cite{Chuang1997, Poyatos1997}. To obtain the process matrix corresponding to a given pulse, we simulate the final state after the pulse with respect to initial states $\ket{0}$, $\ket{1}$, $(\ket{0}+i\ket{1})/\sqrt{2}$, and $(\ket{0}-\ket{1})/\sqrt{2}$. The simulations are done by numerically solving the master equation in the Lindblad form~\cite{gardiner2004quantum}:
\be
\dot \rho(t) = -i[H(t), \rho(t)]+\sum_n\frac{\kappa_n}{2} (2\hat{L}_n\rho\hat{L}_n^\dagger - \hat{L}_n^\dagger\hat{L}_n\rho -\rho\hat{L}_n^\dagger\hat{L}_n), \label{eq:gates:mastereq}
\ee
where $\hat{L}_n$ and $\kappa_n$ are the Lindblad operator and the decoherence rate for a certain decoherence source respectively. Here we consider only one decoherence mechanism, energy relaxation thus we only have one term in Eq.~\eq{gates:mastereq}, with $\hat{L} = \sigma_-$ and $\kappa = 1/T_1$. The actual values of the qubit parameters are chosen to be the same as those characterized in the experiments discussed in Ref.~\onlinecite{Deng2015}, where $\Delta = 2\pi \times 2.288$~GHz and $T_1 = 2$~$\mu$s. We note that the value of $T_1$ here is shorter than those obtained in other experiments with flux qubits, and therefore the fidelities we obtain are conservative estimates. Once we have the final states with respect to the four initial states, we can reconstruct the process matrix $\chi$ using standard QPT. From the process matrix $\chi$ obtained by QPT and its ideal counterpart $\chi_\t{ideal}$, we can directly calculate the process fidelity, defined as $F_p = \t{Tr}[\chi_\t{ideal} \chi]$. The gate fidelity $F_g$, defined as the state fidelity of the process output state with respect to the ideal output state averaged over all possible input states, can be related to the process fidelity $F_p$ by $F_g = (dF_p + 1)/(1 + d)$ with $d$ the dimension of the system~\cite{Nielsen2002249}. We simulate the gate fidelity of operations $R_x(\theta)$ and $R_y(\theta)$ implemented by pulses with $\phi = 0$ and $-\pi/2$ at $A_m = \Delta/4$. In these simulations, we choose the rise times of the pulses $t_r = 1/\omega$ and $2.6/\omega$ for $R_x(\theta)$ and $R_y(\theta)$ respectively for the maximum suppression of the nonadiabatic transitions on the rising edge. In Fig.~\ref{fig:pulsefidelity}(c) and (d), we plot the gate fidelities versus the maximum amplitude duration $t_p$ with different fall times. For short fall times ($t_f \sim 1/\omega$), the gate fidelity oscillates as a function of $t_p$. This is a signature of coherent nonadiabatic transitions between the Floquet modes. The highest-fidelities are attained when the combinations of $t_p$ and $t_f$ lead to small nonadiabatic transitions (minima in Fig.~\ref{fig:pulsefidelity}(a) and (b)). For longer fall times ($t_f \gtrsim 5/\omega$), the ramping of the falling edge is approximately adiabatic thus the gate fidelity has suppressed oscillations and is limited by $T_1$.

\begin{figure}[]
  \includegraphics[width=80mm]{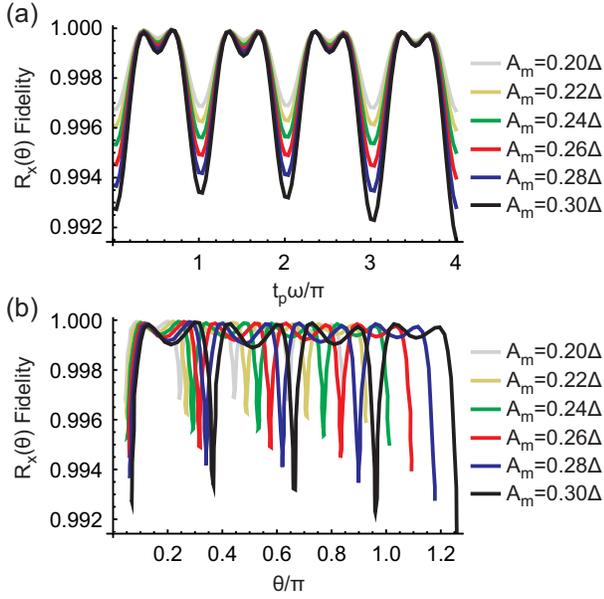}
  \caption{Simulated gate fidelities of the implemented $R_x(\theta)$ versus the maximum amplitude duration $t_p$ (a) or the rotation angle $\theta$ given by $\theta = \int_0^tdt' \Delta\epsilon(t')$ (b). The pulses are defined with $\phi = 0$, $t_r = 1/\omega$ and $t_f = 1/\omega$ and various values of $t_p$ and $A_m$. While the optimal $t_p$ for the coherent suppression of nonadiabatic transitions is independent of $A_m$, different pulses with the same $t_r$, $t_f$, and $t_p$ can have different rotation angles $\theta$ depending on $A_m$.} \label{fig:fidelity_vs_A}
\end{figure}
So far, we have shown that high-fidelity operations $R_x(\theta)$ and $R_y(\theta)$ can be achieved for some specific rotation angle $\theta$ which depends on the optimal combination of $t_r$, $t_p$, and $t_f$. Performing univeral quantum computation, however, requires engineering pulses that correspond to arbitrary rotation angles. This goal is easy to achieve in our optimization method. As discussed in Sec.~\ref{sec:nonadiabatic}, the optimal $t_r$ and $t_f$ are almost independent of the maximum driving amplitude $A_m$ as long as $A_m \lesssim \Delta$. In Fig.~\ref{fig:fidelity_vs_A}(a), we plot the gate fidelity of $R_x(\theta)$ versus $t_p$ and $A_m$ with the optimal $t_r = 1/\omega$ and a short $t_f = 1/\omega$. We show that the optimal $t_p$, which minimizes nonadiabatic transitions and leads to high-fidelity operations, is also independent of $A_m$. In Fig.~\ref{fig:fidelity_vs_A}(b), we plot the gate fidelity of $R_x(\theta)$ versus $\theta$ at several values of $A_m$. We show that by fixing $t_r$, $t_p$, and $t_f$ to their the optimal values and sweeping $A_m$, high-fidelity rotations for any rotation angles $\theta$ can be achieved. Therefore, to fully optimize a pulse for a given rotation operation, one can first determine $t_r$, $t_p$, and $t_f$ based on the rotation axis and a rough range of $A_m$ and then fine-tune $A_m$ by targeting a precise rotation angle given by $\theta = \int_0^tdt' \Delta\epsilon(t')$. As an example, we optimize pulses for the operations $R_x(\pi/2)$ and $R_y(\pi/2)$ in the single-qubit Clifford group~\cite{Calderbank1997} using qubit parameters detailed in Ref.~\onlinecite{Deng2015}. For $R_x(\pi/2)$, the optimized pulse with $t_r = 1/\omega$, $t_f = 1/\omega$, $t_p = 5.30/\omega$, and $A_m = 0.249\omega$ has a gate fidelity of $0.99983$. For $R_y(\pi/2)$, the optimized pulse has a fidelity of $0.99986$ with $t_r = 2.6/\omega$, $t_f = 1/\omega$, $t_p = 4.27/\omega$, and $A_m = 0.270\omega$. The above fidelities are limited by $T_1$, where we assumed $\omega = \Delta = 2\pi\times 2.288$~GHz and $T_1 = 2$~$\mu s$.

Next, we consider optimization with an experimental limitation, as introduced by the time resolution and bandwidth of the ultrafast arbitrary waveform generator used in Ref.~\onlinecite{Deng2015}. We perform pulse optimization by first rounding the calculated optimal $t_r$, $t_f$, and $t_p$ to the 40~ps time resolution of the waveform and limiting the shortest $t_{r}$ and $t_{f}$ to 80~ps. We then optimize $A_m$ to obtain the highest possible gate fidelity. We obtain a fidelity of $0.99962$ for $R_x(\pi/2)$ with the pulse parameters: $t_r = 80$~ps, $t_f = 80$~ps, $t_p = 360$~ps, and $A_m = 2\pi\times 0.5674$~GHz. For $R_y(\pi/2)$, we obtain a fidelity of $0.99972$ with $t_r = 200$~ps, $t_f = 80$~ps, $t_p = 280$~ps, and $A_m = 2\pi\times 0.6223$~GHz. Enforcing a finite time resolution reduces the effectiveness of the suppression of nonadiabatic transitions; however, the resulting fidelities are only slightly lower than those with unconstrained time resolution.

In practice, the achievable pulses are limited by electronics and experimental setup, and the actual pulse times therefore deviate from the designed values. For moderate timing constraints corresponding to a realistic superconducting qubit setup, the suppression of nonadiabatic transitions is still acceptable and the rotation angle error can be compensated by further optimizing the pulse amplitude $A_m$ which can be set with high resolution using standard equipment. Combined with quantum process characterization techniques such as randomized benchmarking~\cite{Kelly2014}, our optimization procedure can be potentially carried out in-situ and therefore is robust against timing uncertainties in practical applications. We note that, thanks to the cosine-function shape of the pulse envelope during the rising and falling edges, this pulse scheme requires a bandwidth of only $2\omega$. For superconducting qubits, which usually have a transition frequency of a few gigahertz, such pulses can be readily generated by a high-bandwidth arbitrary waveform generator.

\section{Comparison with other pulse optimization methods} \label{sec:OQC}
In this section, we discuss the connection between FIESTA and other methods in optimal quantum control (OQC). For an in depth discussion of quantum control, we refer the reader to the review articles~\cite{Brif2010, glaser_training_2015,koch_controlling_2016}. The objective in OQC is maximizing the fidelity of a quantum state or of a quantum operation, given a Hamiltonian with controllable parameters. Constraints on the parameters can be applied, including amplitude constraints and time-domain constraints. Time-domain constraints are of two types: frequency bandwidth constraints or parametrization in terms of analytical functions. A research topic that is closely related to OQC is the quantum speed limit, addressing operations implemented with high fidelity in the minimum possible amount of time~\cite{caneva_optimal_2009,ashhab_speed_2012,Hegerfeldt2013}.

Gradient ascent pulse engineering (GRAPE)~\cite{khaneja_optimal_2005} and Krotov~\cite{somloi_controlled_1993} optimal quantum control rely on assuming a piecewise-constant form of the control parameters. In these methods, an iterative algorithm is used to steadily increase the fidelity and eventually converge to optimal values of the control parameters in all the time intervals. With sufficient discretization, continuously varying control parameters can be approximated with good accuracy. Other control methods, including chopped random-basis quantum optimization (CRAB)~\cite{caneva_chopped_2011} and Gradient Optimization of Analytic conTrols (GOAT)~\cite{machnes2015gradient} assume the Hamiltonian to be expressed in terms of analytic functions, with a set of parameters entering these functions to be optimized. Derivative Removal by Adiabatic Gate (DRAG)~\cite{motzoi2009simple} is concerned with the reduction of state leakage in a weakly anharmonic system, providing an approximate analytical solution to this problem.

A problem discussed extensively in literature is the control of a qubit with a transverse driving term limited in absolute value. It was shown that the optimal solutions for single-qubit rotations consist of sudden switching of the control between its extreme values, and the optimal time is related to the inverse qubit transition frequency (see Ref.~\onlinecite{Boscain2006}, \onlinecite{Garon2013}, and also \onlinecite{hirose_time-optimal_2015} and references therein).

In contrast to these approaches, closed-loop optimal control is a method in which control is optimized in a system with imperfect knowledge of the system Hamiltonian or control transfer functions, based on feedback from a fidelity measurement performed on the physical system~\cite{egger_adaptive_2014}.

FIESTA is used to perform optimization of quantum gates with resonant driving and a simple pulse shape characterized by three times (pulse rise, top, and fall) and an amplitude (the maximum pulse amplitude). This particular optimization problem has elements related to amplitude constraints and bandwidth constraints explored with other methods. One important characteristic of this pulse shape is that it is motivated by experimental implementation constraints, where using a frequency band around the qubit transition frequency can be conveniently implemented using analog quadrature modulation. Optimization with amplitude constraints can be done using GRAPE or analytically in some cases. Bandwidth constraints can be implemented in GOAT~\cite{machnes2015gradient} or approximately, using filtered optimal pulses~\cite{hirose_time-optimal_2015}. Our approach is physically motivated by the observation that even in the strong driving regime the dynamics can be simply described in terms of three unitaries, for the rise, central, and fall parts of the pulses. We showed that optimizing only the four involved parameters ($t_{r}$, $t_{p}$, $t_{f}$, and $A_{m}$) extremely high fidelities can be attained, with a total pulse duration of the order of $2\pi/\Delta$, thus approaching the quantum speed limit~\cite{Hegerfeldt2013}. We expect that the optimization of these parameters is also suitable for a closed-loop approach, and that high fidelities can be attained with moderate levels of errors arising from the transfer function frequency dependence and qubit detuning.

We note that Bartels and Mintert~\cite{bartels_smooth_2013} discussed a numerical optimization method for control with short pulses based on Floquet theory. The discussion in Ref.~\onlinecite{bartels_smooth_2013} focuses on the implementation of single-qubit gates, the method is applicable straightforwardly to optimizing single-qubit gates. Nevertheless, we expect in general that optimized pulses obtained using their method will have rather complicated envelopes, possibly complicating experimental implementations.

We find that in the strong driving regime the optimal times for $x$ and $y$ rotations obtained with FIESTA are different from each other, and these optimal times depend on the maximum amplitude $A_{m}$. These results bear an interesting connection with the discussion presented in Ref.~\onlinecite{hirose_time-optimal_2015} (see also references therein) of optimized pulses with amplitude constraints and no bandwidth constraints.

We finally comment on the relevance of our work for physical implementations of optimal quantum control. Superconducting flux qubits, for which we have recently observed Floquet dynamics~\cite{Deng2015}, are an ideal two-level system to apply these results, because higher energy levels are well separated from the lowest two levels, even when driving field amplitudes are larger than $\Delta$. NV centers have also been used in strong driving time domain experiments by Fuchs et al.~\cite{Fuchs:2009ca}. In a recent experiment Scheuer et al.~\cite{scheuer_precise_2014} demonstrated the implementation of optimal control in the strong driving regime using optimization based on CRAB. We also expect our theoretical framework to be extendible to optimization of strong driving control for multi-level systems, with superconducting qubits being a case of particular interest.

\section{Conclusion} \label{sec:conclusion}
We have analyzed the dynamics of a two-level system under strong resonant pulses from the perspective of (non)adiabatic evolutions in the Floquet picture. We presented derivations of the approximate analytical expressions for the quasienergies and Floquet states of the system under consideration. We analyzed the effects of pulse shaping using the adiabatic theory in the Floquet picture. We have shown that when the driving amplitude of a pulse varies slowly, the system remains in its initial superposition of the Floquet states and the dynamics are governed by a dynamic phase that depends on the evolution of the quasienergies over time. This phase corresponds to a qubit state rotation, generalizing the notion of Rabi oscillations to the case of large driving amplitudes. We have also analyzed and quantified the nonadiabatic transitions between the Floquet states by employing the adiabatic perturbation theory as well as exact numerical simulations. We found that the first order adiabatic perturbation theory agrees very well with the exact numerics when the driving amplitude is comparable with the qubit frequency and breaks down at very high driving amplitude. In addition, we found that with suitable pulse shaping the nonadiabatic transitions can be coherently suppressed. Finally, we presented FIESTA, an optimization scheme for the pulse shapes that minimize the nonadiabatic transitions. We shown that high-fidelity single-qubit operations can be achieved in very short times, significantly alleviating the effect of decoherence. Furthermore, our optimized control scheme requires pulses with very simple shapes and is robust against experimental limitations and parameter uncertainties. These pulses can be achieved with presently available arbitrary-waveform generators and is therefore ready to be implemented in experiments using superconducting qubits. Our method is an addition to the toolset of quantum control, providing a simple recipe for high fidelity control of a single qubit in the strong driving regime. We expect that the theoretical framework established here can be generalized to optimize the control of multi-level quantum systems.

\begin{acknowledgments}
We thank Frank Wilhelm for useful discussions. We acknowledge support from NSERC, Canada Foundation for Innovation, Ontario Ministry of Research and Innovation, Industry Canada. During this work, AL was supported by an Early Research Award.
\end{acknowledgments}

\appendix*
\section{Derivation of the analytical forms of quasienergies and Floquet modes} \label{appendix}
This appendix contains the derivation of the analytical forms of the quasienergies and Floquet modes in Sec.~\ref{sec:Floquet}.
\begin{widetext}
Starting from the Floquet Hamiltonian given by Eq.~\eq{FloquetHamiltonian}, we now perform a basis transformation where the basis states after the transformation are related to those before the transformation by the formula
\begin{eqnarray}
\ket{\tilde{u}_{j,n,0}} & = & \{\dots , J_{-1}\left(\frac{A}{\omega}\right) , 0 , J_0\left(\frac{A}{\omega}\right) , 0 , J_1\left(\frac{A}{\omega}\right) , 0 , \dots \} , \nonumber \\
\ket{\tilde{u}_{j,n,1}} & = & \{\dots , 0 , J_{-1}\left(-\frac{A}{\omega}\right) , 0 , J_0\left(-\frac{A}{\omega}\right) , 0 , J_1\left(-\frac{A}{\omega}\right) , \dots \}.
\end{eqnarray}
In other words, when expressed in the original, time-domain representation,
\ba
\ket{\tilde{u}_{j,n,0}} & = \sum_{m=-\infty}^{\infty} e^{im\omega t} J_{m-n} \left( \frac{A}{\omega} \right) \ket{0}, \nn \\
\ket{\tilde{u}_{j,n,1}} & = \sum_{m=-\infty}^{\infty} e^{im\omega t} J_{m-n} \left( -\frac{A}{\omega} \right) \ket{1}.
\ea
This transformation is motivated by the fact that it diagonalizes the Hamiltonian $H_F$ when $\Delta=0$. When considering strong driving, one can use the point $\Delta=0$ as a starting point and treat $\Delta$ as a small parameter. The Hamiltonian in the new basis reads
\begin{equation}
H_F = \left(
\begin{array}{cccccccc}
\ddots & & & & & & & \\
& (n-1) \omega & {-\frac{\Delta}{2}J_0\left(\frac{2A}{\omega}\right)} & 0 & {-\frac{\Delta}{2}J_1\left(\frac{2A}{\omega}\right)} & 0 & {-\frac{\Delta}{2}J_2\left(\frac{2A}{\omega}\right)} & \\
& {-\frac{\Delta}{2}J_0\left(\frac{2A}{\omega}\right)} & (n-1) \omega & {\frac{\Delta}{2}J_1\left(\frac{2A}{\omega}\right)} & 0 & {-\frac{\Delta}{2}J_2\left(\frac{2A}{\omega}\right)} & 0 & \\
& 0 & {\frac{\Delta}{2}J_1\left(\frac{2A}{\omega}\right)} & n \omega & {-\frac{\Delta}{2}J_0\left(\frac{2A}{\omega}\right)} & 0 & {-\frac{\Delta}{2}J_1\left(\frac{2A}{\omega}\right)} & \\
& {-\frac{\Delta}{2}J_1\left(\frac{2A}{\omega}\right)} & 0 & {-\frac{\Delta}{2}J_0\left(\frac{2A}{\omega}\right)} & n \omega & {\frac{\Delta}{2}J_1\left(\frac{2A}{\omega}\right)} & 0 & \\
& 0 & {-\frac{\Delta}{2}J_2\left(\frac{2A}{\omega}\right)} & 0 & {\frac{\Delta}{2}J_1\left(\frac{2A}{\omega}\right)} & (n+1) \omega & {-\frac{\Delta}{2}J_0\left(\frac{2A}{\omega}\right)} & \\
& {-\frac{\Delta}{2}J_2\left(\frac{2A}{\omega}\right)} & 0 & {-\frac{\Delta}{2}J_1\left(\frac{2A}{\omega}\right)} & 0 & {-\frac{\Delta}{2}J_0\left(\frac{2A}{\omega}\right)} & (n+1) \omega & \\
& & & & & & & \ddots
\end{array}
\right).
\end{equation}
A sufficiently large truncated version of the matrix $H_F$ yields eigenvectors and eigenvalues that converge well. A size of $100 \times 100$ for example is easily sufficient for reliable numerical calculations. As explained in Ref.~\onlinecite{Deng2015}, truncating this Hamiltonian to a $2 \times 2$ matrix gives good results in the strong-driving limit, but it fails in the weak driving limit. As also explained in Ref.~\onlinecite{Deng2015}, a truncation that allows an analytical solution while giving a good approximation for the quasienergies for both weak and strong driving is obtained by using the Hamiltonian
\begin{equation}
H_{F,4 \times 4} = \left(
\begin{array}{cccc}
- \omega & {-\frac{\Delta}{2}J_0\left(\frac{2A}{\omega}\right)} & 0 & {-\frac{\Delta}{2}J_1\left(\frac{2A}{\omega}\right)} \\
{-\frac{\Delta}{2}J_0\left(\frac{2A}{\omega}\right)} & - \omega & {\frac{\Delta}{2}J_1\left(\frac{2A}{\omega}\right)} & 0 \\
0 & {\frac{\Delta}{2}J_1\left(\frac{2A}{\omega}\right)} & 0 & {-\frac{\Delta}{2}J_0\left(\frac{2A}{\omega}\right)} \\
{-\frac{\Delta}{2}J_1\left(\frac{2A}{\omega}\right)} & 0 & {-\frac{\Delta}{2}J_0\left(\frac{2A}{\omega}\right)} & 0
\end{array}
\right).
\label{eq:app:4x4TruncatedHamiltonian}
\end{equation}
Inspired by the symmetry in the above matrix, which is most easily seen in the limiting case $A=0$, we now perform a basis transformation $\tilde{H}_{F,4\times 4} = S^{\dagger} H_{F,4 \times 4} S$, with
\begin{equation}
S = \frac{1}{\sqrt{2}} \left(
\begin{array}{cccc}
1 & 1 & 0 & 0 \\
1 & -1 & 0 & 0 \\
0 & 0 & 1 & 1 \\
0 & 0 & 1 & -1
\end{array}
\right).
\end{equation}
We obtain
\begin{equation}
\tilde{H}_{F,4\times 4} = \frac{1}{2}
\left(
\begin{array}{cccc}
-2\omega-{\Delta J_0\left(\frac{2A}{\omega}\right)} & 0 & 0 & {\Delta J_1\left(\frac{2A}{\omega}\right)} \\
0 & -2\omega+{\Delta J_0\left(\frac{2A}{\omega}\right)} & {-\Delta J_1\left(\frac{2A}{\omega}\right)} & 0 \\
0 & {-\Delta J_1\left(\frac{2A}{\omega}\right)} & -{\Delta J_0\left(\frac{2A}{\omega}\right)} & 0 \\
{\Delta J_1\left(\frac{2A}{\omega}\right)} & 0 & 0 & {\Delta J_0\left(\frac{2A}{\omega}\right)}
\end{array}
\right).
\end{equation}
\end{widetext}
This matrix can be split into two $2 \times 2$ decoupled blocks. The outer block gives quasienergies that are close to (or even degenerate with) quasienergies that were ignored when we truncated the infinite-dimensional matrix $H_F$ to a $4 \times 4$ matrix. We can therefore expect that the quasienergies obtained from the outer block will not be accurate, and indeed they are not. We therefore focus on the inner block:
\begin{equation}
\frac{1}{2}
\left(
\begin{array}{cccc}
-2\omega + {\Delta J_0\left(\frac{2A}{\omega}\right)} & {-\Delta J_1\left(\frac{2A}{\omega}\right)} \\
{-\Delta J_1\left(\frac{2A}{\omega}\right)} & {-\Delta J_0\left(\frac{2A}{\omega}\right)}
\end{array}
\right).
\label{eq:app:2x2TruncatedHamiltonian}
\end{equation}

The eigenvalues or quasienergies are given by:
\ba
\epsilon_0 & = \frac{1}{2} \left( - \omega - \sqrt{ \left[ \omega - \Delta J_0\left(\frac{2A}{\omega}\right) \right]^2 + \Delta^2 J_1^2\left(\frac{2A}{\omega}\right) } \right), \nonumber \\
\epsilon_1 & = \frac{1}{2} \left( - \omega + \sqrt{ \left[ \omega - \Delta J_0\left(\frac{2A}{\omega}\right) \right]^2 + \Delta^2 J_1^2\left(\frac{2A}{\omega}\right) } \right). \label{eq:app:QuasienergiesAnalytical}
\ea

The Floquet modes are given by the corresponding eigenvectors of the Floquet Hamiltonian. In the basis of Eq.~\eq{app:2x2TruncatedHamiltonian}, these eigenvectors are given by
\begin{equation}
\ket{\tilde{u}_0} = \left( \begin{array}{c} {\cos\frac{\theta}{2}} \\ \\ {\sin\frac{\theta}{2}} \end{array} \right) \;\; , \;\; \ket{\tilde{u}_1} = \left( \begin{array}{c} {-\sin\frac{\theta}{2}} \\ \\ {\cos\frac{\theta}{2}} \end{array} \right),
\end{equation}
where
\begin{equation}
\tan\theta = \frac{\Delta J_1\left(\frac{2A}{\omega}\right)}{\omega - \Delta J_0\left(\frac{2A}{\omega}\right)}. \label{eq:app:theta}
\end{equation}
We now note that we have made two transformations to go from the Floquet Hamiltonian $H_F$ in Eq.~(\ref{eq:FloquetHamiltonian}) to $\tilde{H}_{F,4\times 4}$, and we need to transform the eigenvectors back to the original basis. In the basis of Eq.~\eq{app:4x4TruncatedHamiltonian}, the Floquet modes are given by
\begin{equation}
\ket{\tilde{u}_0} = \frac{1}{\sqrt{2}} \left( \begin{array}{c} {\cos\frac{\theta}{2}} \\ \\ {-\cos\frac{\theta}{2}} \\ \\ {\sin\frac{\theta}{2}} \\ \\ {\sin\frac{\theta}{2}} \end{array} \right) \;\; , \;\; \ket{\tilde{u}_1} = \frac{1}{\sqrt{2}} \left( \begin{array}{c} {-\sin\frac{\theta}{2}} \\ \\ {\sin\frac{\theta}{2}} \\ \\ {\cos\frac{\theta}{2}} \\ \\ {\cos\frac{\theta}{2}} \end{array} \right).
\end{equation}
Transforming further to the basis of Eq.~\eq{Hamrotated}, we find the expressions
\ba
\ket{u_{0,n}} &= \frac{1}{\sqrt{2}}\left( \begin{array}{c} {\cos\frac{\theta}{2}J_{n+1}\left(\frac{A}{\omega}\right) + \sin\frac{\theta}{2}J_{n}\left(\frac{A}{\omega}\right)} \\ \\ {-\cos\frac{\theta}{2}J_{n+1}\left(-\frac{A}{\omega}\right) + \sin\frac{\theta}{2}J_{n}\left(-\frac{A}{\omega}\right)} \end{array} \right), \nonumber \\
\ket{u_{1,n}} &= \frac{1}{\sqrt{2}}\left( \begin{array}{c} {-\sin\frac{\theta}{2}J_{n+1}\left(\frac{A}{\omega}\right) + \cos\frac{\theta}{2}J_{n}\left(\frac{A}{\omega}\right)} \\ \\ {\sin\frac{\theta}{2}J_{n+1}\left(-\frac{A}{\omega}\right) + \cos\frac{\theta}{2}J_{n}\left(-\frac{A}{\omega}\right)} \end{array} \right). \label{eq:app:PeriodicPartsOfFloquetStatesOriginalBasis}
\ea

%

\end{document}